\documentclass[a4paper,11pt]{article}
\usepackage{pos}

\usepackage{graphicx}
\usepackage{subcaption}
\def\beq{\begin{equation}}   
	\def\eeq{\end{equation}}
\def\bea{\begin{eqnarray}}  
	\def\eea{\end{eqnarray}} 
\def\nn{\nonumber}
\def\eps{\epsilon}

\def\doubletilde#1{\widetilde{\vphantom{\raise 1.5pt \hbox{#1}}\smash{\kern -2pt\widetilde{#1}}}}

\def\e{\epsilon}
\def\d{\hbox{d}}

\def\nn{\nonumber}

\def\e{\epsilon}
\def\d{\hbox{d}}

\def\nn{\nonumber}

\newcommand{\code}[1]{\textsc{#1}}

\title{N$^3$LO Antenna Functions for Final-State Radiation}

\author*[a]{Matteo Marcoli}

\affiliation{
	$^a$Physik-Institut, Universit\"at Z\"urich, Winterthurerstrasse 190, 8057 Z\"urich, Switzerland\\
}

\emailAdd{matteo.marcoli@physik.uzh.ch}

\abstract{We discuss recent progress in the calculation of integrated antenna functions for final-state radiation at N$^3$LO in QCD. Antenna functions are directly extracted from physical matrix elements for the decay of a colour-singlet state into partons. In order to compute their integrated counterparts, which are necessary ingredients for fixed-order calculations in QCD, each allowed final state has to be individually integrated over the inclusive phase space of the unresolved radiation. We analytically integrated two-, three-, four- and five-particle final-state matrix elements for the decay of a virtual photon and a Higgs boson into partons, considering up to third-order corrections in the strong coupling constant $\alpha_s$. The integration over the inclusive phase space is performed relying on well-established reverse unitarity and IBP reduction techniques. We conclude with preliminary comments about the structure of N$^3$LO subtraction terms in the context of the antenna subtraction method.}

\FullConference{16th International Symposium on Radiative Corrections: Applications of Quantum Field Theory to Phenomenology (
RADCOR2023)\\
28th May - 2nd June, 2023\\
Crieff, Scotland, UK\\}

\renewcommand{\hookAfterAbstract}{%
	\par\bigskip
	\textsc{ArXiv ePrints}:
	\href{https://arxiv.org/abs/2211.08446}{2211.08446}, \href{https://arxiv.org/abs/2304.11180}{2304.11180}
}

\begin{document}
\maketitle

\section{Introduction}

The increasing accuracy of experimental measurements and the future High-Luminosity phase of the Large Hadron Collider demand extremely precise phenomenological predictions. NNLO fixed-order calculations in QCD are nowadays available for a large variety of processes, but computations at an even higher order are needed to meet the required precision for benchmark processes and observables. In the past decade, several N$^3$LO-accurate inclusive and differential calculations appeared, to the point that the list of the associated publications is too sizeable to be included in this proceeding. However, phenomenological predictions at N$^3$LO are only available for simple processes with $2\to 1$ kinematics: Drell-Yan, Deep Inelastic Scattering (DIS), Higgs production (VBF and $gg$-induced), possibly with decay to bottom quark, two Higgs bosons (inclusive) or an Higgs boson and a vector boson (inclusive), $e^+ e^- \to jj$ and $e^+ e^- \to t\bar{t}$ (inclusive). The techniques employed for the calculation of differential predictions are $q_T$-slicing~\cite{Catani:2007vq} and projection-to-Born~\cite{Cacciari:2015jma}. The application of these methods to N$^3$LO calculations is at the moment restricted to the aforementioned class of processes, namely colour-singlet production and associated crossings. Moreover, the \mbox{$q_T$-slicing} approach is particularly challenging on the computational side, since it relies on non-local cancellations of infrared singularities.

In parallel to the phenomenological applications, remarkable efforts have been made to identify the singular behaviour of QCD amplitudes at N$^3$LO and all the results for the description of triple-unresolved limits at tree-level, double-unresolved limits at one-loop and single-unresolved limits at two-loops are nowadays available. See~\cite{Jakubcik:2022zdi} for a list of references. Such calculations are the starting point for the formulation of a future local subtraction scheme at N$^3$LO. 

Among subtraction schemes for NNLO calculations, antenna subtraction~\cite{Gehrmann-DeRidder:2005btv,Currie:2013vh,Chen:2022ktf} has proven to be sufficiently flexible and general to be applied for the calculation of the NNLO correction to numerous processes. The core ingredients of this method are the \textit{antenna functions}~\cite{Gehrmann-DeRidder:2005btv}: universal objects which can be used to describe the singular behaviour of QCD matrix elements in infrared limits. Antenna functions are derived from physical matrix elements for the decay of colour-singlet states into partons~\cite{Gehrmann-DeRidder:2004ttg,Gehrmann-DeRidder:2005alt,Gehrmann-DeRidder:2005svg}. We indicate a generic $\ell$-loop matrix element for the decay into $n$ partons as $M_n^{\ell}$. Such matrix elements can be analytically integrated over the inclusive phase space of the unresolved radiation to obtain \textit{integrated antenna functions}, which are necessary for the definition of the antenna subtraction scheme.

The purpose of the work presented in this talk is the extension of the NNLO results of~\cite{Gehrmann-DeRidder:2004ttg,Gehrmann-DeRidder:2005alt} to N$^3$LO. At this perturbative order, the total decay rate for a colour-singlet state is given by
\begin{equation}\label{eqn:matelems}
	\sigma^{(3)} = \int \d\Phi_5\,M_{5}^0
	+ \int \d\Phi_4\,M_{4}^1
	+ \int \d\Phi_3\,M_{3}^2
	+ \int \d\Phi_2\,M_{2}^3\,,
\end{equation}
where $\d\Phi_n$ indicates the $n$-particle phases space. In the following, we briefly illustrate how we computed each term in~\eqref{eqn:matelems} to obtain integrated quark-antiquark and gluon-gluon antenna functions.

\section{Method}\label{sec:method}

We consider the the two decays $\gamma^*\to q\bar{q}$ and $H \to gg$, used to extract N$^3$LO quark-antiquark and gluon-gluon antenna functions respectively. For the decay of a Higgs boson into gluons, we consider the heavy-top effective theory, with the QCD
Lagrangian supplemented with an effective Lagrangian given by
\begin{equation}
	\mathcal{L}_{gg} = - \frac{\lambda_0}{4} H G_a^{\mu\nu} G_{a,\mu\nu}\,,
\end{equation}
with $G_a^{\mu\nu}$ the renormalised gluon field-strength, $H$ the Higgs field
and $\lambda_0$ the bare effective coupling~\cite{Chetyrkin:1997iv,Kniehl:1995tn,Chetyrkin:1997un}. We work in dimensional regularization, with the customary number of dimensions $d=4-2\eps$. The renormalization of the loop correction is performed in the
$\overline{\text{MS}}$ scheme. We replace the bare coupling $\alpha_0$ with the
renormalised coupling $\alpha_s$ according to
\begin{eqnarray}\label{eq:alfaren}
	\alpha_0\,\mu_0^{2\e}\,S_\e &=& \alpha_s\,\mu^{2\e}\Bigg[
	1- \frac{\beta_0}{\e}\left(\frac{\alpha_s}{2\pi}\right) 
	+\left(\dfrac{\beta_0^2}{\e^2}-\dfrac{\beta_1}{2\e}\right)\left(\frac{\alpha_s}{2\pi}\right)^2 
	\nonumber \\ && \phantom{\alpha_s\,\mu^{2\e}\Bigg[} 
	- \left(\dfrac{\beta_0^3}{\e^3}-\dfrac{7}{6}\dfrac{\beta_1\beta_0}{\e^2}
	+ \dfrac{1}{3}\dfrac{\beta_2}{\e}\right)   \left(\frac{\alpha_s}{2\pi}\right)^3
	+{\cal O}(\alpha_s^3) \Bigg]\,,
\end{eqnarray}
where $S_\e = (4\pi)^\e e^{-\e\gamma}$, $\alpha_0$ is the bare coupling and the $\beta_i$ coefficients are given for example in~\cite{Gehrmann:2010ue}. We fix the renormalisation scale $\mu^2$ to be the invariant mass of the decaying particle $q^2$. We also renormalize the
effective coupling $\lambda_0=Z_{\lambda}\lambda$~\cite{Spiridonov:1988md}
as described in~\cite{Gehrmann:2010ue}.

\begin{figure}[t]
	\centering
	\begin{subfigure}[b]{0.22\linewidth}
		\includegraphics[width=\linewidth]{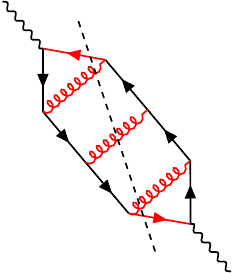}
		\caption{5 cut-propagators}
		\label{fig:a}
	\end{subfigure}
	\begin{subfigure}[b]{0.22\linewidth}
		\includegraphics[width=\linewidth]{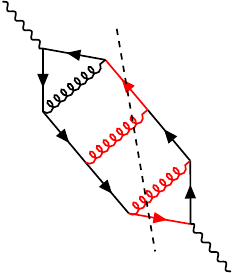}
		\caption{4 cut-propagators}
		\label{fig:b}
	\end{subfigure}
	\begin{subfigure}[b]{0.22\linewidth}
		\includegraphics[width=\linewidth]{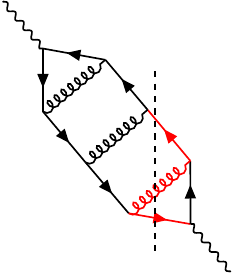}
		\caption{3 cut-propagators}
		\label{fig:c}
	\end{subfigure}
	\begin{subfigure}[b]{0.22\linewidth}
		\includegraphics[width=\linewidth]{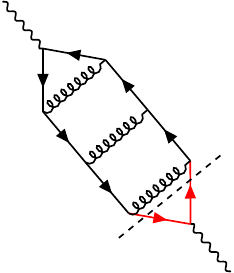}
		\caption{2 cut-propagators}
		\label{fig:d}
	\end{subfigure}
	\caption{Example of a four-loop massless photon self-energy with different possible choices of cut propagators (in red).}\label{fig:cuts}
\end{figure}

Phase-space integrals are related to loop integrals by the reverse unitarity
relation~\cite{Cutkosky:1960sp,Anastasiou:2002yz} which reformulates the on-shellness condition for a final-state massless momentum $p$ as an inverse propagator \textit{on cut}:
\begin{equation}
	2\pi i\delta^+(p^2) = \frac{1}{p^2 - i0} - \frac{1}{p^2 + i0}\,.
\end{equation}
This way, all the integrals
present in the matrix elements in \eqref{eqn:matelems} can be expressed as cuts of the four-loop correction to the propagator of the colourless particle and we can leverage techniques developed for the computation of loop integrals. The relevant cuts of a particular diagram are depicted in Fig.~\ref{fig:cuts} in the case of photon decay.

Our method is explained in the following. At first, four-loop diagrams with two external legs are generated with \code{QGRAF}~\cite{Nogueira:1991ex} using a model which includes the Standard Model QCD particles and couplings, as well as a set of fields for cut-propagators, which are also allowed to couple to regular particles. Most arrangements of the cut-fields are unphysical. In fact, a diagram only contributes to the physical integrated cross section if the cuts divide it into exactly two connected graphs, each attached to an external current. Moreover, a specific number of cuts and loops is required for each layer of the calculation. Finally, only cuts which fulfil momentum conservation and do not contain self-energy insertions on cut-propagators are retained. We subsequently insert the Feynman rules into the selected diagrams, compute the colour and Dirac algebra and impose on-shellness conditions for cut propagators. Keeping track of which propagators are cut in any diagram, we can define for every auxiliary topology of the original four-loop propagator a set of \textit{cut integral families} which cover all the integrals appearing in the matrix elements. The integrals appearing in the matrix elements have up to eleven propagators in the denominator and a maximum of four (photon decay) or five (Higgs decay) scalar products in the numerator, in line with the calculation of the three-loop quark and gluon form factors in \cite{Gehrmann:2010ue}. For each layer, we use \code{Reduze2}~\cite{vonManteuffel:2012np} to reduce all integrals to a set of master integrals, finding $22$, $27$, $35$ and $31$ master integrals for the two-, \mbox{three-}, four-, and five-particle final state respectively. The master integrals required for the N$^3$LO  calculation were computed in~\cite{Gituliar:2018bcr,Magerya:2019cvz}. \code{FORM}~\cite{Vermaseren:2000nd,Kuipers:2012rf} together with \code{Mathematica} and \code{Python} scripts were used extensively throughout the calculation.


\section{Results}

The results were presented explicitly in~\cite{Jakubcik:2022zdi} and~\cite{Chen:2023fba}, where we report the analytical expressions for the inclusive integration of all the individual partonic sub-channels at NLO, NNLO and N$^3$LO, up to transcendental weight 6, normalized with respect to the corresponding LO inclusive decay width. As an example, we report here the result for the leading-colour contribution to the three-gluon two-quark final state for the photon decay:
\begin{eqnarray}
	&&{\cal T}^{(\gamma,3)}_{q\bar{q}ggg}\Big|_{N^2} = 
	+\frac{1}{\e^6}\left(\frac{1}{2}\right)+\frac{1}{\e^5}\left(\frac{331}{108}\right) +\frac{1}{\e^4}\left(\frac{11843}{648}-\frac{31}{24}\pi^2\right)\nn \\
	&&+\hspace{0.2cm}\frac{1}{\e^3}\left(\frac{259867}{2592}-\frac{10745}{1296}\pi^2-\frac{439}{18}\zeta_{3}\right)\nn \\
	&&+\hspace{0.2cm}\frac{1}{\e^2}\left(\frac{6302057}{11664}-\frac{394223}{7776}\pi^2-\frac{6239}{36}\zeta_{3}+\frac{21853}{25920}\pi^4\right)\nn \\
	&&+\hspace{0.2cm}\frac{1}{\e}\Bigg(\frac{815913157}{279936}-\frac{26347837}{93312}\pi^2-\frac{181151}{162}\zeta_{3}+\frac{75767}{17280}\pi^4+\frac{13993}{216}\pi^2\zeta_{3}-\frac{10946}{45}\zeta_{5}\Bigg)\nn \\
	&&+\hspace{0.2cm}\frac{736904809}{46656}-\frac{107045579}{69984}\pi^2-\frac{49920557}{7776}\zeta_{3}+\frac{7130357}{311040}\pi^4\nn \\
	&&+\hspace{0.2cm}\frac{67895}{144}\pi^2\zeta_{3}-\frac{103894}{45}\zeta_{5}-\frac{93257}{1306368}\pi^6+\frac{7861}{12}\zeta_{3}^2 + \mathcal{O}(\e).
\end{eqnarray}

We perform several checks on our calculation. First of all, we ensure that we recover the known quark and gluon three-loop form factors~\cite{Baikov:2009bg,Gehrmann:2010ue} from our two-parton final-state matrix elements. Secondly, the sum of the renormalized and fully integrated contributions in~\eqref{eqn:matelems} has to be free from any infrared singularity. Finally, we can validate the finite part, checking that it agrees with the known total decay rate at N$^3$LO~\cite{Gorishnii:1990vf,Surguladze:1990tg,Baikov:2005rw,Moch:2007tx}.

The obtained results give us the the opportunity to inspect the infrared singularity structure of different multiplicity final states when the integration over unresolved radiation is performed. Focusing on the most subleading-colour contribution within the perturbative corrections to the decay of a virtual photon, we can interpret the deepest singularities as the exponentiation of multiple photon-like (abelian) emissions from the hard quarks. In particular, considering a matrix element with $r$ abelian real emissions and $\ell=k-r$ abelian virtual emissions, after analytical integration over the loop momenta and the inclusive phase space of the radiation, the infrared poles satisfy:
\begin{eqnarray}
	{\cal T}^{q\bar{q},(k)}_{{\cal I}}\Big|_{\rm abelian}
	&=& 
	{\cal N}\binom{k}{r}(-1)^{r} \left(I^{(1)}_{q\bar{q}}\right)^k
	+ \mathcal{O}(\e^{-2k+2})\,,
	\label{eq:binomial}
\end{eqnarray}
where $I^{(1)}_{q\bar{q}}$ is the one-loop infrared insertion operator defined in~\cite{Catani:1998bh} for a quark-antiquark dipole, and ${\cal N}$ an appropriate normalization. 

An interesting contribution appearing for the first time at N$^3$LO is the singlet term in the decay of a virtual photon, which comes with the characteristic charge and colour factor
\begin{equation}
	\left(\sum_q e_q\right)^2 d^{abc}d^{abc}=\left(\sum_q e_q\right)^2 \left(N-\dfrac{4}{N}\right),
\end{equation}
where $e_q$ represents the electric charge of the quark $q$ and $d^{abc}d^{abc}$ is the quartic Casimir of SU($3$). We observe that such contribution is infrared-finite for the two- and three-parton final states, while it exhibits a $\eps^{-1}$ pole in the four- and five-parton final states. We can relate the origin of such singularity to antisymmetric terms in the one-lop triple collinear splitting functions~\cite{Catani:2003vu} and in the tripole term in the current for the emission of a soft gluon and a soft quark-antiquark pair~\cite{Catani:2022hkb,DelDuca:2022noh}.

In~\cite{Chen:2023fba} we compare the infrared structure of each layer of~\eqref{eqn:matelems} for the decay of a virtual photon with the decay of an Higgs boson to a massless quark-antiquark pair, for which we consider a non-vanishing Yukawa coupling. We observe substantial agreement among the two processes in the coefficients of the deepest infrared poles, as well as in the highest-weight transcendental constants for every power of $\eps$.

In general, the complete analysis of the infrared singularities of each contribution in~\eqref{eqn:matelems} at N$^3$LO is particularly involved. In~\cite{Jakubcik:2022zdi,Chen:2023fba} we provide various comments and observations, but a general explanation is possible only with a better understanding of the unresolved structures at the integrated level.

\section{A sketch of the N$^3$LO subtraction infrastructure}\label{sec:sub}

In this section we present a very preliminary study of the structure of the subtraction terms within a future extension of the antenna subtraction method at N$^3$LO. At this perturbative order there are triple-real (RRR), double-real-virtual (VRR), double-virtual-real (VVR) and triple-virtual (VVV) corrections. It is possible to identify the required ingredients for each layer counting the number of loops and additional real emissions in suitable combinations of antenna functions and matrix elements. We denote an $\ell$-loop $n$-parton antenna function with $X_n^{\ell}$, and its integrated counterpart with $\mathcal{X}_n^{\ell}$. The new ingredients we have recently computed and discussed in this talk are therefore $\mathcal{X}_5^{0}$, $\mathcal{X}_4^{1}$ and $\mathcal{X}_3^{2}$. We can now consider a process with $h$ external hard partons, generically described by the LO matrix element ${\cal M}_h^0$. The structure of the subtraction terms for the four layers of an N$^3$LO calculation would roughly look like:
\begin{eqnarray}
	RRR:&& \hspace{-0.5cm}{\color{red}+X_3^0 {\cal M}_{h+2}^0 }{\color{blue}+\big(X_4^0 - X_3^0 X_3^0\big){\cal M}_{h+1}^0} {\color{black}+\big(X_5^0 - X_4^0 X_3^0 + X_3^0 X_3^0 X_3^0\big){\cal M}_{h}^0}  \\	
	&&\vspace{0.3cm}	\nonumber \\
	VRR:&&\hspace{-0.5cm} {\color{red}+\mathcal{X}_3^0 {\cal M}_{h+2}^0} {\color{blue}+ \big(X_3^1+ \mathcal{X}_3^0 X_3^0\big) {\cal M}_{h+1}^0+ X_3^0{\cal M}_{h+1}^1}{\color{black}+ \big(X_4^0+ X_3^0 X_3^0\big) {\cal M}_h^1} \nonumber \\				
	&&\hspace{-0.5cm} + \big(\mathcal{X}_3^0 X_3^0 X_3^0+ \mathcal{X}_3^0 X_4^0 + X_4^1 - X_3^1 X_3^0\big) {\cal M}_h^0  \\
	&&\vspace{0.3cm}	\nonumber \\				
	VVR:&& \hspace{-0.5cm}{\color{blue}+ \big(\mathcal{X}_4^0 + \mathcal{X}_3^1 + \mathcal{X}_3^0 \mathcal{X}_3^0\big)\ {\cal M}_{h+1}^0 + \mathcal{X}_3^0 {\cal M}_{h+1}^1}{\color{black}+ X_3^2 {\cal M}_h^0- X_3^1 {\cal M}_h^1} \nonumber \\				
	&&\hspace{-0.5cm}- X_3^0 {\cal M}_h^2+ \mathcal{X}_3^0 X_3^0 {\cal M}_h^1+ \mathcal{X}_3^0 X_3^1 {\cal M}_h^0+ \mathcal{X}_4^0 X_3^0 {\cal M}_h^0  \nonumber \\				
	&&\hspace{-0.5cm}+ \mathcal{X}_3^1 X_3^0 {\cal M}_h^0+ \mathcal{X}_3^0 X_3^0 X_3^0 {\cal M}_h^0  \\		
	&&\vspace{0.3cm}	\nonumber \\		
	VVV:&& \hspace{-0.5cm}+ \big(\mathcal{X}_5^0 + \mathcal{X}_4^1 + \mathcal{X}_3^2+ \mathcal{X}_4^0 \mathcal{X}_3^0+ \mathcal{X}_3^1 \mathcal{X}_3^0+ \mathcal{X}_3^0 \mathcal{X}_3^0 \mathcal{X}_3^0\big)\ {\cal M}_h^0 \nonumber \\				
	&&\hspace{-0.5cm}+\big(\mathcal{X}_4^0+\mathcal{X}_3^1+\mathcal{X}_3^0 \mathcal{X}_3^0\big)\ {\cal M}_h^1+ \mathcal{X}_3^0 {\cal M}_h^2 
\end{eqnarray}
The formulae above should be interpreted as a list of structures we expect to appear in each layer, rather then exact expressions for the subtraction terms. Relative different signs among terms indicate the removal of singular behaviour, which would be otherwise over-subtracted by the whole subtraction term. Terms in red and blue respectively represent the NLO and NNLO sub-sectors of the subtraction infrastructure. One can expect that, more than the employment of genuine N$^3$LO antenna functions, the arrangement of lower-orders ingredients will pose a highly non-trivial task, due to the proliferation of allowed combinations of NLO and NNLO antenna functions and matrix elements.

\section{Conclusions and outlook}

We presented the analytical integration of individual partonic channels for the decay of a virtual photon and a Higgs boson to partons, up to five final-state particles and perturbative order $\alpha_s^3$. The two-loop three-parton, one-loop four-parton and tree-level five-parton cases are directly related to N$^3$LO quark-antiquark and gluon-gluon antenna functions for final-state radiation. The results we present can therefore be regarded as a first necessary step towards the extension of the antenna subtraction approach at N$^3$LO. 

We plan to perform an analogous calculation~\cite{Chen:2023egx} for the decay of a neutralino into a gluino and a gluon~\cite{Gehrmann-DeRidder:2005svg}, which allows for the extraction of quark-gluon antenna functions. With this, we will obtain a complete set of integrated antenna functions for final-state radiation at N$^3$LO. We envisage future work to assemble a consistent set of subtraction terms for the removal of infrared singularities within each individual layer of an N$^3$LO calculation, along the lines of the preliminary discussion presented here in section~\ref{sec:sub}.

\bibliographystyle{JHEP}
\bibliography{bib_N3LO_ant}

\end{document}